\documentclass[9pt,twocolumn,twoside]{osajnl}

\journal{ol} % Choose journal (ao, aop, josaa, josab, ol)

\setboolean{shortarticle}{true}
\ifthenelse{\boolean{shortarticle}}{\colorlet{color2}{color2b}}{\colorlet{color2}{color2}} % Automatically switches colors for short articles

\title{Absolute sensitivity of phase measurement in an SU(1,1) type
	interferometer}

\author[1]{Wei Du}
\author[1]{Jun Jia}
\author[1,3]{J. F. Chen}
\author[4,*]{Z. Y. OU}
\author[2,3,*]{Weiping Zhang}

\affil[1]{Quantum Institute of Light and Atoms,Department of Physics \\East China Normal University,Shanghai,200241,People's Republic of China\\}
\affil[2]{Department of Physics and Astronomy,Tsung-Dao Lee Institute, Shanghai Jiao Tong University \\Shanghai 200240, People's Republic of China\\}
\affil[3]{Collaborative Innovation Center of Extreme Optics Shanxi University, Taiyuan, Shanxi 030006, People's Republic of China\\}
\affil[4]{Department of Physics Indiana University-Purdue University Indianapolis, 402 North Blackford Street Indianapolis, Indiana 46202, USA\\}
\affil[*]{Corresponding author: zou@iupui.edu, wpz@sjtu.edu.cn}

\dates{Compiled \today}

\ociscodes{(190.4420) Nonlinear optics; (270.0270) Quantum optics; (020.0020) Atomic and molecular physics; (270.5565) Quantum communications; (270.5585) Quantum information and processing.}

\doi{\url{http://dx.doi.org/10.1364/ao.XX.XXXXXX}}

\begin{abstract}
Absolute sensitivity is measured for the phase measurement in an SU(1,1) type interferometer and the results are compared to that of a Mach-Zehnder interferometer operated under the condition of the same intra-interferometer intensity. The interferometer is phase locked to a point with the largest quantum noise cancellation, and a simulated phase modulation is added in one arm of SU(1,1) interferometer. Both the signal and noise level are estimated at the same frequency range, and we obtain 3 dB improvement in sensitivity for the SU(1,1) interferometer over the Mach-Zehnder interferometer. Our results demonstrate a direct phase estimation, and may pave the way for practical applications of nonlinear interferometer.
\end{abstract}

\setboolean{displaycopyright}{true}
\begin{document}

\maketitle
\thispagestyle{fancy}

\ifthenelse{\boolean{shortarticle}}{\ifthenelse{\boolean{singlecolumn}}{\abscontentformatted}{\abscontent}}{}

Optical phase estimation is indispensable in the process of measurement, because it can help us determine how small the phase we can detect\cite{Degen,Huntington}. For example, it has been used for phase-sensing which can measure an already known phase with small interval\cite{Pooser,Nagata}. It can also be utilized to track an unknown phase\cite{Yonezawa}. Optical interferometer is one of the most important device in phase measurement for its wide application in modern metrology\cite{Abadie}. With the strong motivation of exploring gravitational wave, e.g., the Laser Interferometer Gravitational-Wave Observatory(LIGO) project\cite{Abbott}, laser interferometer has been rapidly developing as one of the most precise device all over the world. On the other hand, the fundamental principles of operation in optical interferometers had never changed ever since its discovery more than 200 years ago, that is, they all rely on linear passive beam splitters for wave splitting and recombination. It has been proven that quantum vacuum noise sets the so-called standard quantum limit(SQL)\cite{Helstrom,Giovannetti} for the sensitivity of this type of interferometers, in which the smallest detectable phase change $\Delta\phi$ requires high laser power running in the interferometer\cite{Abbott1}. However, high laser power induces problems like strong backaction of mirrors\cite{Sheon}, and restrict the further improvement of phase sensitivity. Squeezed states can be utilized for the suppression of the vacuum noise and break the standard quantum limit for improvement of sensitivity\cite{Xiao}.

Quantum correlations between the optical fields in the interferometer is proved to enhance the phase sensitivity beyond SQL, to Heisenberg limit $\Delta\phi\sim1/N$. Recently, a new type of interferometer was realized that relies on nonlinear optical processes for beam spitting and recombination. The so-called SU(1,1) interferometer(SUI) was first proposed by Yurke et al\cite{Yurke} about 30 years ago and became experimentally implementable\cite{jing} after the discovery of a four-wave mixing(FWM) process in hot atomic vapor\cite{McCormick} which serves as the nonlinear beam splitter. The twin beams have been utilized to study a large amount of quantum physics experiment, such as quantum entangled images\cite{Boyer}, tunable delay of continuous EPR entanglement state\cite{Marino1}, quantum plasmonic sensing\cite{Fan}, low noise amplifier\cite{PD Lett,Kong,Corzo}, and so on. In particular, our group has experimentally investigated such an SU(1,1) type interferometer whose beam splitting and recombination elements are replaced by FWM process in the hot rubidium atomic ensemble\cite{jing,F}. With the signal of the SU(1,1) interferometer amplified by the nonlinear FWM process and the noise nearly maintained, we have successfully observed a 4dB enhancement of signal-to-noise(SNR) ratio compared to conventional linear Mech-Zehnder interferometer. The improvement of the measured signal is the DC signal\cite{Lukens}, which not match to the quantum noise reduction bandwidth. Also, a direct phase sensitivity estimation has never been implemented in this SU(1,1) interferometer.

\begin{figure*}[tbp]
	\centering
	\includegraphics[height=9.5cm,width=18cm]{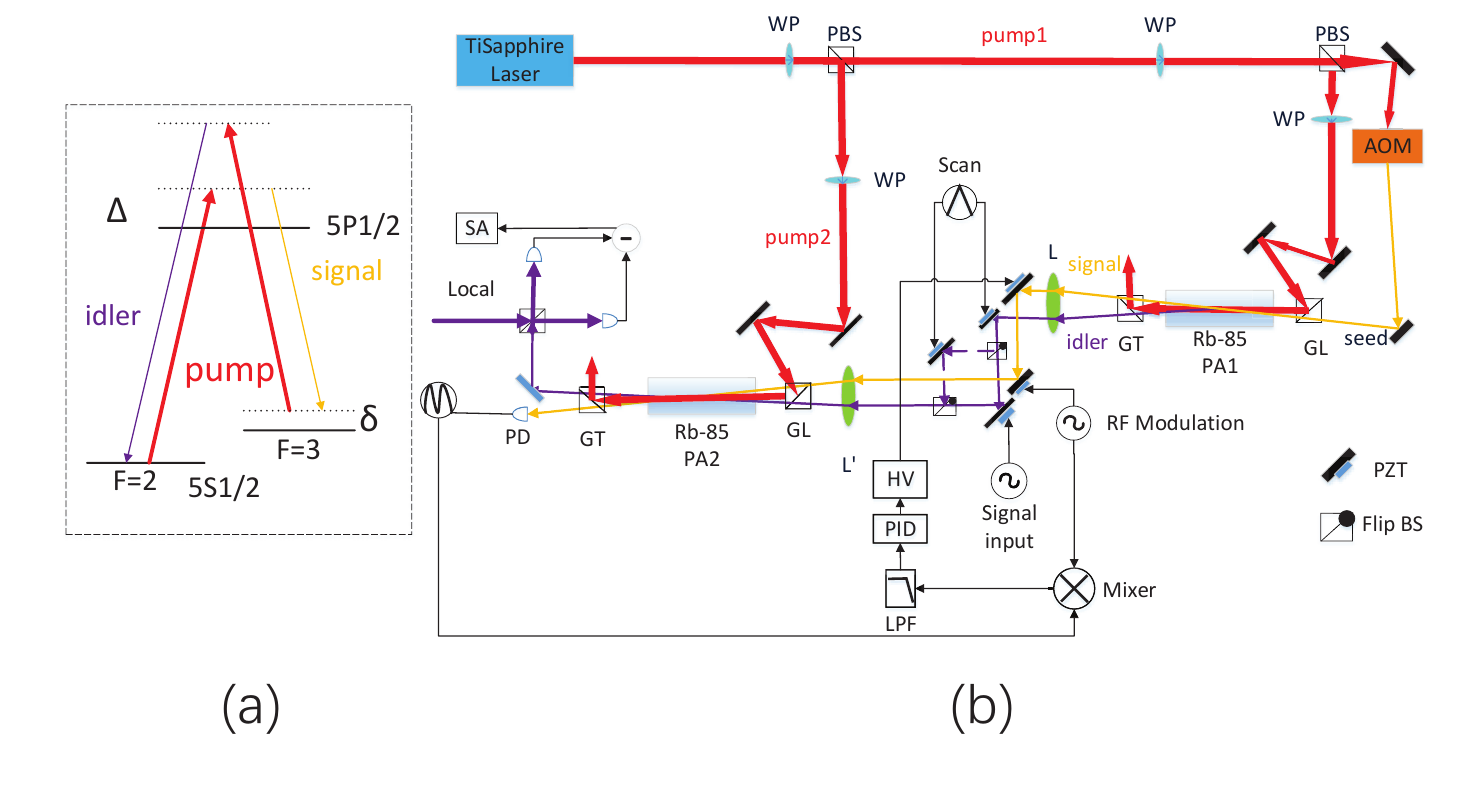}% Here is how to import EPS art
	\caption{Experimental layout for absolute sensitivity of phase measurement. (a) a Double-$ \lambda $ system at the D1 line of $^{85}$Rb. (b) Detailed experiment arrangement. WP: wave plate; PBS: polarization beam splitters; AOM: acousto-optic modulator; GL: Glan-Laser polarizer; GT: Glan-Thompson polarizer; PD: photodiode; SA: spectum analyzer; LPF: low pass filter; HV:high voltage amplifier. A traditional MZ interferometer is set in the SU(1,1) interferometer by flip BS, so the SU(1,1) interferometer and the MZ interferometer can test the same phase change which is modulated by PZT.}
	\label{fig:setup}
\end{figure*}

In this letter, we report a direct phase estimation of a SU(1,1) interferometer. For nonlinear interferometer, the quantum noise cancellation usually occurs at dark port\cite{McClelland}, so a locking system is designed to lock to the dark fringe of the SU(1,1) interferometer. Consequently, we measure noise level at the dark fringe. At the low noise region around 1.6 MHz, a weak phase modulation signal is injected into one arm of both Cl and NCl interferometer. We measure for the first time the strength of signal and noise at the same range of frequency, and estimate the absolute value of phase change. We therefore verify that our SU(1,1) interferometer can detect a signal half as small as conventional interferometer.

The detailed arrangement of our experiment is shown in Fig.1. Our SU(1,1) interferometer use two parametric amplifier(PA1,PA2) for splitting and
combination, which are based on non-degenerate four-wave-mixing process.In theory,the FWM process PA1 can be described as

\begin{eqnarray}
\hat{a}_{s}=\sqrt{G}\hat{a}_{s}^{in}+\sqrt{g}\hat{a}_{i}^{in\dag},\hat{a}_{i}=\sqrt{G}\hat{a}_{i}^{in}+\sqrt{g}\hat{a}_{s}^{in\dag}
\end{eqnarray}

Where $G$ is the amplitude factor of our parametric amplification, and $G-g=1$.In our system $a_{s}$ is coherent state, $a_{i}$ is vacuum state. When we inject a phase shift $\varphi $ on the idler field,and we can get the output fields of interferometer on idler port:

\begin{eqnarray}
\hat{a}_{i}^{out}=(\sqrt{G_{2}g_{1}}e^{i\varphi}+\sqrt{G_{1}g_{2}})\hat{a}_{in}^{\dag}
\end{eqnarray}

For simplification,we assume $G1=G2=G$ and $g1=g2=g$. So we can obtain the output intensity in idler port of non-conventional interferometer:

\begin{eqnarray}
I_{NCI}&=&<\hat{a}_{i}^{out\dag}\hat{a}_{i}^{out}>=2G^{2}g^{2}(\alpha^{2}+1)(1+\cos\varphi)\\
&=&2G^{2}I_{ps}^{NCL}(1+\cos\varphi)\nonumber
\end{eqnarray}

Where $I_{ps}^{NCL}=g^{2}(\alpha^{2}+1)$ is the intensity of phase-sensing field on the arm of idler.

For comparison, we give the output intensity of conventional MZ interferometer with the same phase:

\begin{eqnarray}
I_{CI}= \frac{1}{2}I_{ps}^{CL}(1+\cos\varphi)
\end{eqnarray}

For the case that the splitting component of MZ interferometer is 50:50 beam splitter, we set $I_{ps}^{CL}=2I_{ps}^{NCL}$ for fair comparison.

As a result,we can find the relation between output of conventional interferometer and Non-conventional interferometer:

\begin{eqnarray}
I_{NCI}=\frac{1}{2G}I_{CI}
\end{eqnarray}

In our previous work\cite{Ou}, we have analyse the noise performance of SU(1,1) interferometer, and demonstrate that the SU(1,1) interferometer have the same noise level as MZ interferometer at dark port,i.e., the vacuum noise level. Combine the amplification of signal as we analyse above,we obtain:

\begin{eqnarray}
\delta_{min}^{NCI}=\frac{1}{\sqrt{2G}}\delta_{min}^{CI}=\frac{1}{\sqrt{2GN}}
\end{eqnarray}

Our experimental setup is shown in Fig.\ref{fig:setup}. The primary light source is a Ti:sapphire laser(Spectra-Physics) whose frequency is locked to a stable reference cavity with a line width of about 30 kHz. This laser can supply the strong pump beams of the FWM process in the hot $^{85}$Rb atomic vapor. The $^{85}$Rb atomic vapor cells (12.5 mm long) are heated to 113$^\circ$C. All the faces of the cells are antireflection-coated to achieve a high transmission efficiency above 97\%. Fig.1(a) shows the atomic energy diagram of $^{85}$Rb. The strong pump beam (red) serves as the two pump fields in four-wave mixing, and produces two weak beams labeled as signal and idler at an angle of $0.4^\circ $ with respect to the pump beam to satisfy the phase match condition. The signal and the idler beams are red- and blue-shifted 3.04 GHz respectively from the pump beam. The pump beams are vertically polarized at a maximum power of 400 mW with a waist of 500 $\mu m$ and are blue-detuned about 1 GHz from the D1 line of rubidium ($5S_{1/2}\rightarrow 5P_{1/2}$, 795 nm). A seed beam is injected along the signal field. It is red-detuned about 3.04 GHz from the pump beam by using an acousto-optic modulator (AOM, Brimrose) which is driven by an RF signal generator (Agilent, N9310A) with a double-pass configuration. The seed beam is horizontally polarized with a maximum power of 200 $\mu W$ with a waist of 250 $\mu m$, and is recombined with the pump beam with a Glan-Laser polarizer before entering the atomic cell.  We can control the power of the beams by using polarization beam splitters (PBS) and half-wave plates. After the atomic cell, the injected seed at the signal mode is amplified and is accompanied by the generation of the idler beam through the four wave-mixing process. This one serves as the beam splitter for the injected seed beam. A second identical system is constructed, serving as the beam combiner to complete the SU(1,1) interferometer.

\begin{figure}[tbp]
	\centering
	\includegraphics[height=7cm,width=8cm]{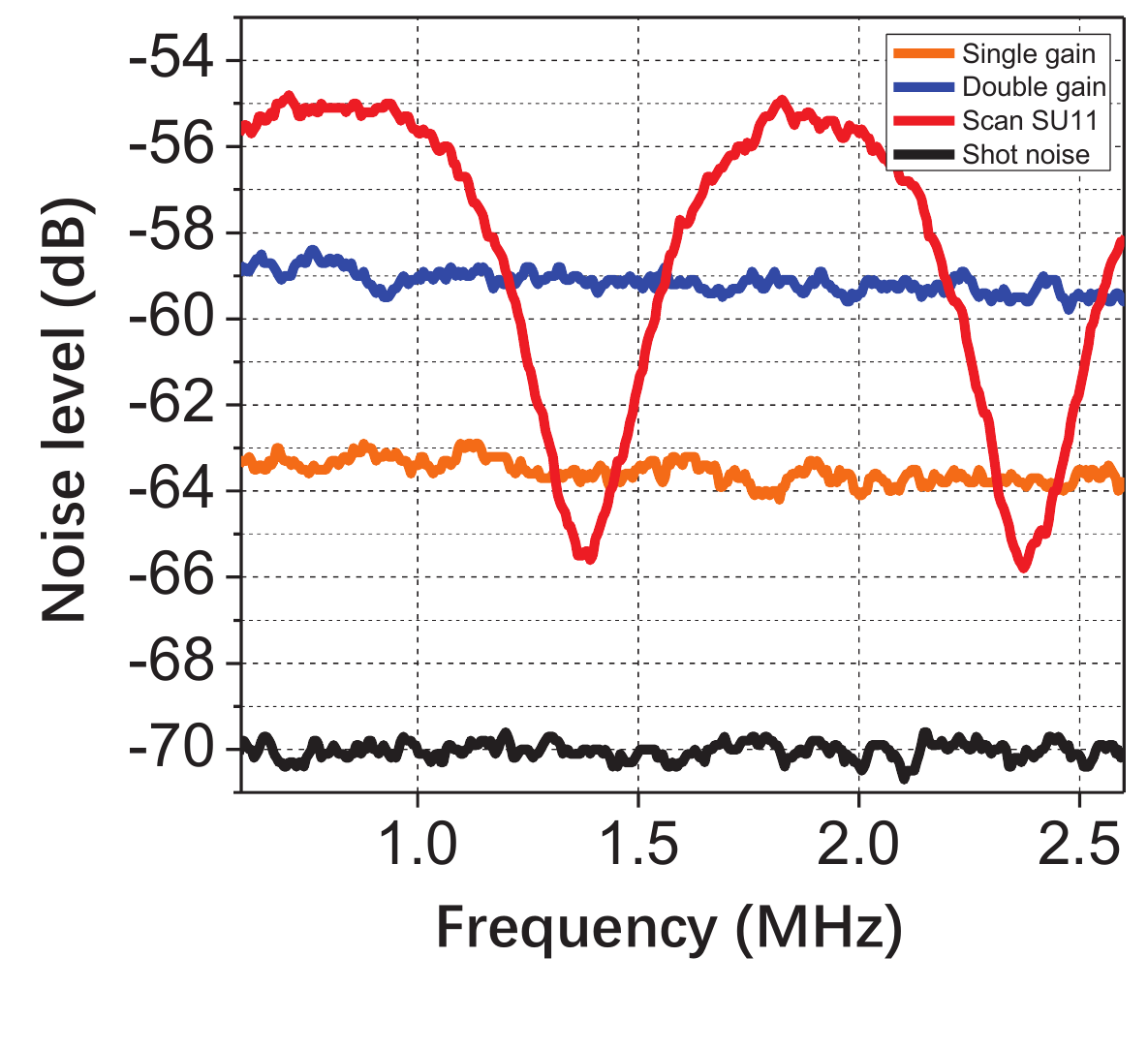}% Here is how to import EPS art
	\caption{The noise level for SU(1,1) interferometer. Measured noise levels from 0.6 to 2.6MHz at the output:the red trace is when the phase $\varphi $ SU(1,1) is scanned; the orange trace is when only PA1 amplify the input light field; the blue trace is when both PA1 and PA2 amplify the input light field; the dark green is the vacuum noise level, also called shot noise, which is the noise level for traditional MZ interferometer. The resolution bandwidth and video bandwidth of the spectrum analyzer is set to be 100 kHz and 100 Hz respectively.}
	\label{fig:noise}
\end{figure}

As shown in Fig.1(b), a 4-F imaging system is used to ensure the mode match between the first and the second cells. Mirrors are mounted on piezoelectric transducers (PZT) to introduce phase change or to lock the interferometer. At the output ports of the second cell, we obtain interference signals in both signal and idler beams with synchronized
phase \cite{jing}.

First, we measure the noise performance of the SU(1,1) interferometer at the idler output port by homodyne detection (HD) and a spectrum analyzer. Fig.\ref{fig:noise} shows the noise spectra from 0.6 MHz to 2.6 MHz under various conditions. The black trace is the shot noise level when we block the input field to HD, which corresponds to the vacuum noise level and is 10dB higher than the electronic noise. The trace in orange shows the noise level of SU(1,1) interferometer when PA1 is operative while no pump is applied to PA2. This is the amplified vacuum noise, which is about 6.2dB higher than the vacuum noise level. The gain of PA1 is G1 = 5. The trace in blue corresponds to the vacuum noise doubly amplified by both PA1 and PA2 in series but we block the idler beam so that no interference occurs. The gain of PA2 is set at G2=3. The most important trace is the red one that represents the noise level of the SU(1,1) interferometer when the phase is scanned via one of the PZTs. The trace shows a phase dependent noise with a minimum at the dark fringe, which will be the working point of the interferometer for phase measurement. In subsequent investigations, we lock the interferometer at this location by a feedback loop as shown in Fig.1.

\begin{figure}[tbp]
	\centering
	\includegraphics[height=9cm,width=8cm]{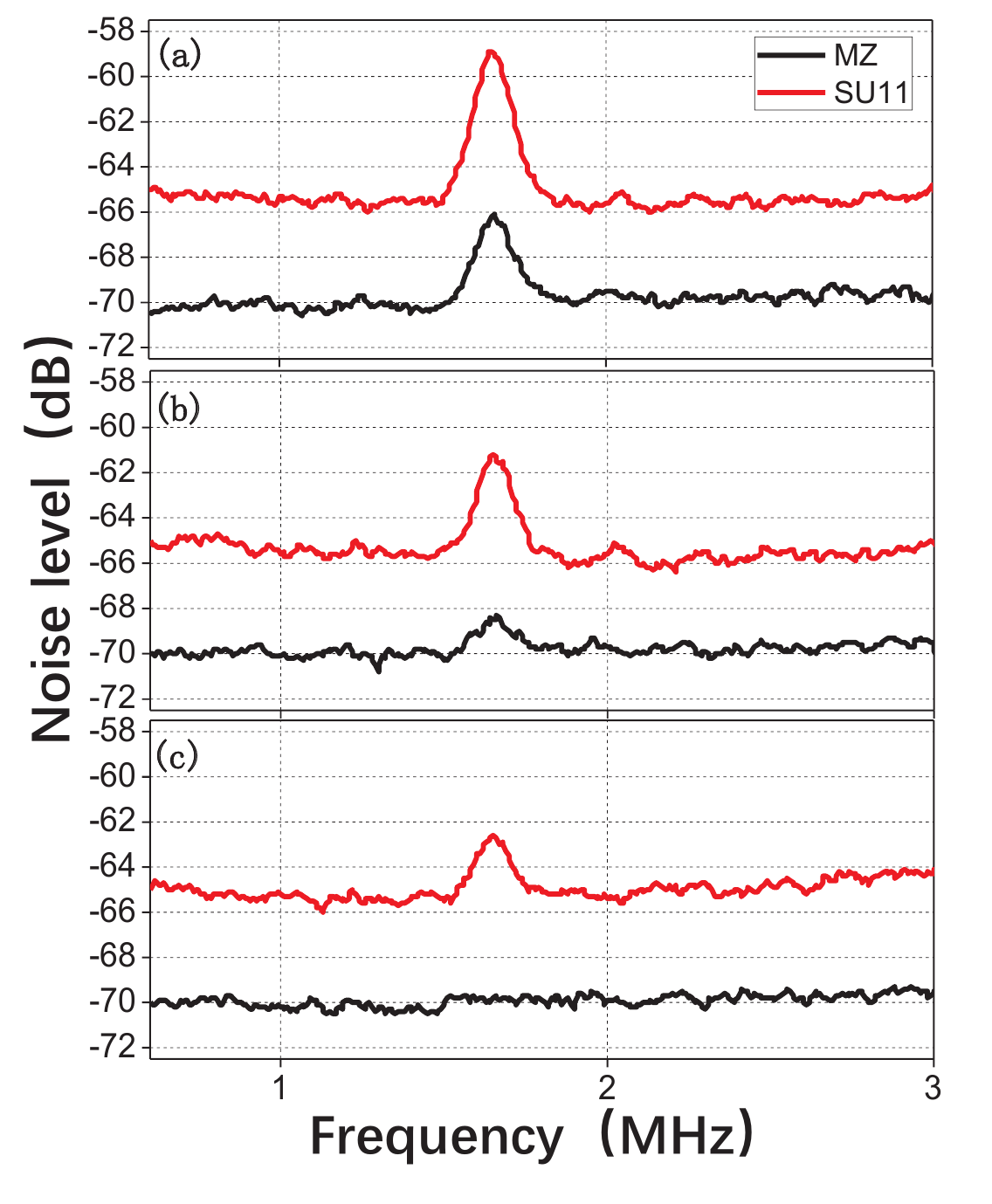}% Here is how to import EPS art
	\caption{The absolute signal level of SU(1,1) and MZ interferometer with different strength of phase modulated signal when $I_{ps}$ is about 5 $\mu W$. (a) Phase change equal to $4.7\times10^{-5}$ rad; (b) Phase change equal to $3.1\times10^{-5}$ rad; (c) Phase change equal to $2.4\times10^{-5}$ rad.}
	\label{fig:phase measurement}
\end{figure}

In our previous study \cite{F} of the SU(1,1) interferometer, we measured the phase signal and the noise independently at different frequencies: phase signal was obtained by a slow scan (100 Hz) but the noise was measured at higher frequency of 1.6 MHz. So, the signal-to-noise ratio (SNR) was implied indirectly by these measurement. Here, we will make a direct measurement of the SNR of phase measurement at the same frequency. To achieve this, we need to have a phase modulation signal at MHz range but the response of PZT is around 10 kHz. However, we notice that the PZT in use has some resonance peaks at around MHz. We select one of these peaks as the modulation frequency. The results of phase measurement are shown in Fig.3 where we show the phase modulation signal and the noise at around 1.6 MHz (red curves) at three different modulation strengths. For comparison, we also show the phase modulation signal measured in a Mach-Zehnder interferometer (MZI), which is formed via two flipping PBS's to redirect the signal beam in between the two PAs. In this way, the MZI measures exactly the same phase modulation signal from the PZT with the same phase-sensing beam (We need to double the injection seed to compensate for the splitting loss of the phase-sensing beam due to the insertion of the PBS). The MZI is also locked at the dark fringe and the noise level is at vacuum noise. The results are shown as the black trace in Fig.\ref{fig:phase measurement}. As can be seen, about 3dB improvement is obtained in SNR for all three settings in the SUI over that in the MZI. Particularly, when the phase modulation is at $\delta = 2.4\times 10^{-5}$ radian, no phase signal is observable with MZI but is still visible with SUI, indicating the superiority of SUI to MZI in phase measurement.

It should be noted that the sizes of the phase modulation are the absolute values of phase change. This is calibrated against an electro-optic modulator of known half voltage placed in the other arm of the MZI. We calibrate the driving voltage to the PZT by observing the same size of phase modulation signal. We find that phase modulations are nonlinearly dependent on the driving voltage at these resonant peaks.

\begin{figure}[tbp]
	\centering
	\includegraphics[height=7cm,width=8cm]{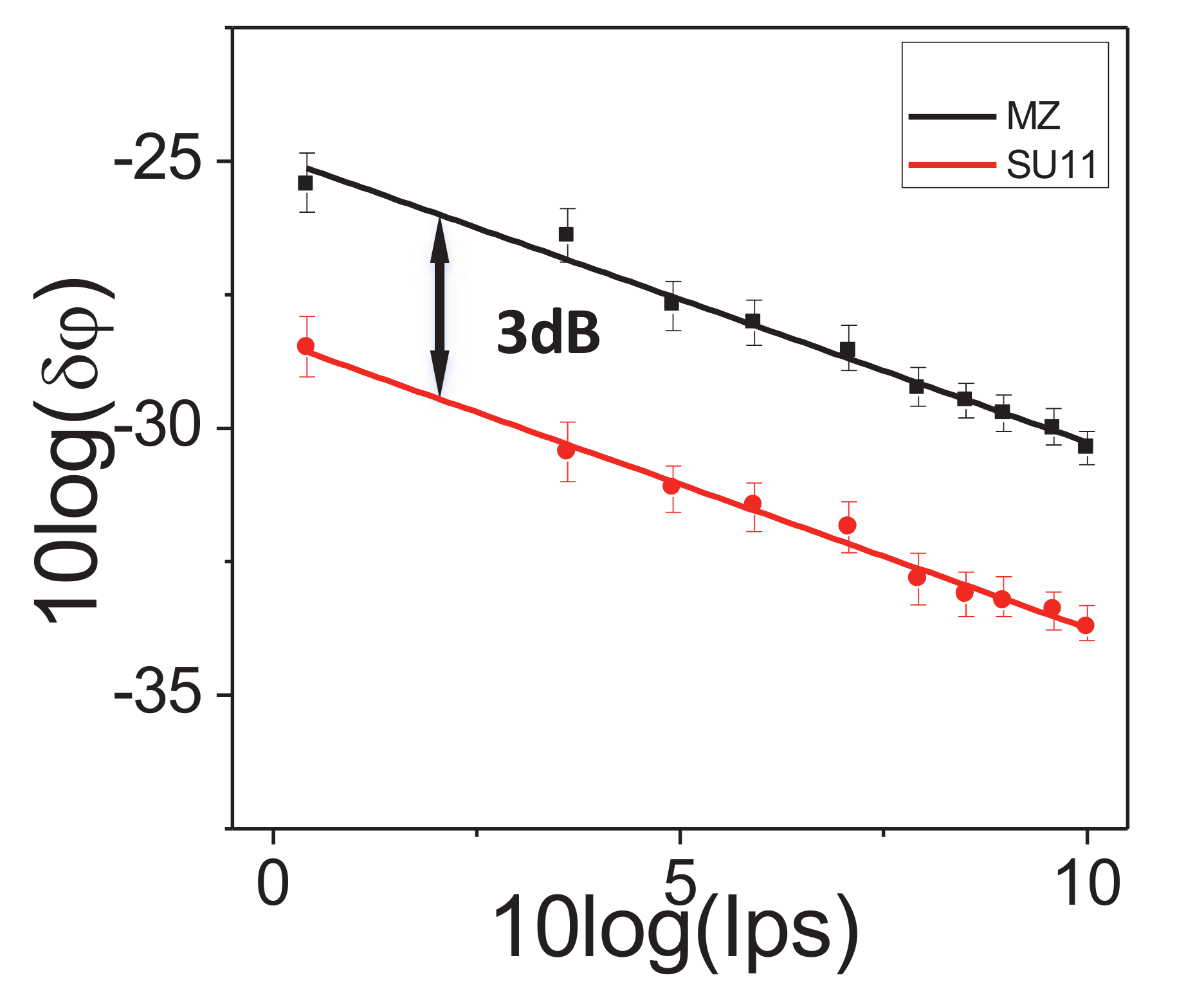}% Here is how to import EPS art
	\caption{The relationship between the $I_{ps}$ and phase for SU(1,1) interferometer and MZ interferometer.}
	\label{fig:compare}
\end{figure}

With calibrated absolute phase measurement, we next measure the minimum observable phase signal under a fixed phase sensing intensity of the probe field. We reduce the driving voltage to the PZT until the SNR is 1 or the peak signal level in Fig.3 is 3dB above the noise level. The corresponding phase is the minimum detectable phase shift under a certain phase sensing intensity. We plot in log-log scale in Fig.4, the measured absolute minimum detectable phases $ \delta\varphi_{min} $ of the SU(1,1) interferometer and those of the MZI as a function of the phase sensing intensity $ I_{ps} $. The error bars are the estimates from the up and down shot-to-shot fluctuations of the signal peak level. Linear fits to the data are obtained with a slope of 0.51$ \pm $0.04 and 0.51$ \pm $0.04 for SUI and MZI, respectively, in consistent with Eq.(6). The difference between the two lines is 3.2 dB,which is the improvement of SUI over MZI.

In conclusion,we have studied the absolute sensitivity of SU(1,1) interferometer in dark port with different kind of parameters,and compare the performance to MZ interferometer. We have directly observed that SU(1,1) interferometer can detect a smaller phase than MZ interferometer under the same operating condition. The improvement is because the SU(1,1) interferometer can amplify the signal but not amplify the noise at the same time. For the case that our interferometer can be operated with very small photon count, which means our interferometer is very helpful for the detection of biological samples, because low light level can avoid the damage of samples\cite{Bowen}. Our technique can be used for phase estimation of nonlinear interferometer, it may find applications in quantum metrology where homodyne measurement is implement for quantum noise detection.

\section*{Funding}

National Key Research and Development
Program of China under Grant No.2016YFA0302001; the National Natural
Science Foundation of China (NSFC) (Grants No.11654005,
11234003, , 11674100, 11129402), and the Science and
Technology Commission of Shanghai Municipality
(STCSM) (Grant No.16DZ2260200).


\begin{thebibliography}{1}

\bibitem{Degen} C. L. Degen, F. Reinhard, and P. Cappellaro, Rev. Mod. Phys. \textbf{89}, 035002 (2017).

\bibitem{Huntington} T. A. Wheatley, D. W. Berry, H. Yonezawa, D. Nakane, H. Arao, D. T. Pope, T. C. Ralph, H. M. Wiseman, A. Furusawa, and E. H. Huntington, Phys. Rev. Lett. \textbf{104}, 093601 (2010).

\bibitem{Pooser} Lawrie, Benjamin, and R. C. Pooser, Optica. \textbf{2}, 393 (2015).

\bibitem{Nagata}T.Nagata, R.Okamoto, J.L.O’Brien, K.Sasaki, S.Takeuchi, Science \textbf{316}, 726(2007).

\bibitem{Yonezawa} H. Yonezawa \textit{et al.,} Science \textbf{337}, 1514 (2012).

\bibitem{Abadie} J. Abadie \textit{et al.,} Nature Physics. \textbf{7}, 962 (2011).

\bibitem{Abbott} B. P. Abbott \textit{et al.,} Phys. Rev. Lett. \textbf{116},061102(2016).

\bibitem{Helstrom}C.W.Helstrom, Quantum Detection and Estimation Theory(Academic Press, New York, 1976).

\bibitem{Giovannetti}V.Giovannetti, S.Lloyd, L.Maccone, Science \textbf{306}, 1330(2004).

\bibitem{Abbott1}B. P. Abbott \textit{et al.,} Science \textbf{256}, 325(1992).

\bibitem{Sheon} Sheon S. Y.Chua, Michael S. Stefszky, Conor M. Mow-Lowry, Ben C. Buchler, Sheila Dwyer, Daniel A. Shaddock, Ping Koy Lam, and David E. McClelland, Opt. Lett. \textbf{36}, 4680 (2011).

\bibitem{Xiao}Min Xiao, L.A. Wu, and H.J. Kimble, Phys. Rev. Lett. \textbf{59}, 278 (1987).   

\bibitem{Yurke} Bernard Yurke, Samuel L.McCall, and John R.Klauder, Phys. Rev. A. \textbf{33}, 4033 (1986).
	
\bibitem{jing}Jietai Jing, Cunjin Liu, Zhifan Zhou, Z. Y. Ou, and Weiping Zhang, Appl. Phys. Lett. \textbf{99}, 011110 (2011).
	
\bibitem{McCormick} C. F. McCormick, A. M. Marino, V. Boyer, and P. D. Lett, Phys. Rev. A. \textbf{78}, 043816 (2008). 	

\bibitem{Boyer}Vincent Boyer, Alberto M.Marino, Raphael C.Pooser, and P.D.Lett, Science \textbf{321}, 544 (2008).

\bibitem{Marino1}A. M. Marino, R. C. Pooser, V. Boyer and P. D. Lett, Nature \textbf{457}, 859 (2009).

\bibitem{Fan}Wenjiang Fan, Benjamin J. Lawrie, and Raphael C.Pooser, Phys. Rev. A. \textbf{92}, 053812 (2015).

\bibitem{PD Lett}RC Pooser, AM Marino, V Boyer, KM Jones, P.D.Lett, Phys. Rev. Lett. \textbf{103}, 010501(2009).

\bibitem{Kong}Jia Kong, F.Hudelist, Z.Y.Ou, and Weiping Zhang, Phys. Rev. Lett. \textbf{111}, 033608(2013).

\bibitem{Corzo}N.V.Corzo, A.M.Marino, K.M.Jones, and P.D.Lett, Phys. Rev. Lett. \textbf{109},043602(2012).
		
\bibitem{F}F.Hudelist, Jia Kong, Cunjin Liu, Jietai Jing, Z. Y. OU and Weiping Zhang, Nat. Commun. \textbf{5}, 3049 (2014).

\bibitem{Lukens}Joseph M. Lukens, Nicholas A. Peters, and Raphael C. Pooser, Opt. Lett. \textbf{41}, 5438 (2016).

\bibitem{McClelland}Kirk McKenzie, Eugeniy Mikhailov, Keisuke Goda, Ping Koy Lam, Nicolai Grosse, Malcolm B.Gray, Nergis Mavalvala and David E.McClelland, Journal of Optics B Quantum and Semiclassical Optics. \textbf{7}, 10(2005).

\bibitem{Ou}Z. Y. Ou, Phys. Rev. A. \textbf{85}, 270 (2012).

\bibitem{Bowen}Michael A. Taylor and Warwick P. Bowen, \textbf{615}, 1 (2016)
\end{thebibliography}
\end{document}